\documentclass{aa} 
\usepackage{graphicx}
\usepackage{natbib}
\bibpunct{(}{)}{;}{a}{}{,}
\usepackage{txfonts}
\begin{document}

   \title{Interpolation of spectra from 3D model atmospheres}
   \author{S. Bertran de Lis \inst{1,} \inst{2}, C. Allende Prieto \inst{1,} \inst{2}, H.-G. Ludwig \inst{3}, \and L. Koesterke \inst{4} }

   \institute{
           Instituto de Astrof\'{\i}sica de Canarias, V\'{\i}a L\'actea, 38205 La Laguna, Tenerife, Spain        
           \and
           Universidad de La Laguna, Departamento de Astrof\'{\i}sica, 38206 La Laguna, Tenerife, Spain
          \and     
          Zentrum f\"ur Astronomie der Universit\"at Heidelberg. Landessternwarte, K\"onigstuhl 12, 69117 Heidelberg, Germany
          \and
          Texas Advanced Computing Center, University of Texas, Austin, TX 78759, USA       
          }

   \date{Received ... / Accepted ...}

   \titlerunning{Interpolation of 3D models of atmospheres}
   \authorrunning{S. Bertran de Lis et al.}

\abstract
   {The use of 3D hydrodynamical simulations of stellar surface convection for model atmospheres is computationally expensive. Although these models have been available for quite some time, their use is limited because of the lack of extensive grids of simulations and associated spectra.}
   {Our goal is to provide a method to interpolate spectra that can be applied to both 1D and 3D models, and implement it in a code available to the community. This tool will enable the routine use of 3D model atmospheres in the analysis of stellar spectra.}
   {We have developed a code that makes use of radial basis functions to interpolate the spectra included in the CIFIST grid of 84 three-dimensional model atmospheres. Spectral synthesis on the hydrodynamical simulations was previously performed with the code ASS$\epsilon$T.}
   {We make a tool for the interpolation of 3D spectra  available to the community. The code provides interpolated spectra and interpolation errors for a given wavelength interval, and a combination of effective temperature, surface gravity, and metallicity. In addition, it optionally provides graphical representations of the RMS and mean ratio between 1D and 3D spectra, and maps of the errors in the interpolated spectra across the parameter space.}
   {}



   \maketitle
%

\section{Introduction}

Spectroscopy is one of the most powerful sources of information for stellar astrophysics. Accurate interpretation of stellar spectra is essential for deriving quantities such as the surface temperature and gravity of stars, their rotation, chemical composition, radial velocity, and so on, and the same is true for galaxy properties from integrated stellar population spectra. The accuracy of the inferred quantities not only depends on the instrumentation we use to acquire the observations, but also on the models adopted to interpret them.

Stellar photospheric spectra are commonly analyzed using classic hydrostatic models of stellar atmospheres, which assume plane-parallel geometry and consider the effect of convective motions in the energetic balance with mixing-length theory \citep{Bohm58}. Despite the success of this approach in reproducing the impact of convection on the temperature gradient, it cannot account for some features observed in real spectra, such as the broadening of spectral lines, their asymmetries, or convective blueshifts. In order to partially compensate for these drawbacks, micro- and macroturbulence parameters should be introduced to fit the spectra.

On the other hand, a new generation of 3D hydrodynamical simulations of stellar atmospheres appeared in the 1980s \citep{Nordlund82}. These models implement a more realistic treatment of convective motions, and thanks to them it is possible to reproduce the asymmetries and convective shifts of spectral lines \citep{Asplund00}. In addition, the real broadening of spectral lines appears naturally without introducing any extra parameters. However, 3D models are not yet extensively used in spectral analysis because of the large computation that they require. 

In the last decade, new grids of 3D models of atmospheres have been produced
for a wide range of stellar parameters. Making use of the CIFIST grid of 3D
simulations \citep{Ludwig09} and the ASS{$\epsilon$}T massively parallel
radiative transfer code \citep{Koesterke08}, we computed a new grid of 3D
spectra \citep[][ and references therein]{Ludwig21}. The wavelength coverage of the spectra spans from 2000 to 30000~\AA, and the parameters of the grid range from 3500 to 7000~K in effective temperature, [Fe/H]\footnote{
We use the standard bracket notation, [a/b] 
$= \log \frac{\rm N(a)}{\rm N(b)}  - 
\log \left(\frac{\rm N(a)}{\rm N(b)}\right)_{\odot}$,
 where N(x) represents the number density of nuclei of the element x.}
from 0 to -3, and $\log g$ from 1.5 to 4.5.

In this paper, we describe the strategy and code that we have developed to interpolate the 3D spectra. Our aim is to provide the means to use synthetic 3D spectra systematically in the analysis of stellar spectra, and to  finally make 3D simulations of stellar atmospheres  accessible and usable.

Section 2 briefly describes the 3D grid of stellar atmospheres, while our
radiative transfer calculations are explained in Section 3. The method and
implementation of the interpolation code are described in Section 4, along with the limitations and approximations made. Section 5 is devoted to explaining
the calculations of errors for the interpolation. Some examples and applications are provided in Section 6. Finally, Section 7 presents a summary of this work.

\section{CIFIST grid of 3D model atmospheres}
\label{models}

The interpolation software presented in this work builds on a set of three-dimensional time-dependent simulations of stellar surface convection, which are part of the CIFIST grid, and were computed using the CO$^{5}$BOLD code \citep{Freytag02,Wedemeyer04,Freytag12}. Convective stellar envelopes, including the stellar atmospheres, are modeled by numerically solving the hydrodynamical equations of mass, momentum, and energy conservation in the presence of a constant gravitational field.
The equation of state takes into account the ionization of H and He, and the formation of H$_{2}$. The simulated stars are assumed to have a solar-scaled chemical composition, adopting solar abundances from \citet{Grevesse98}, with the exception of those for carbon, nitrogen, and oxygen (CNO), which were updated following \citet{Asplund05}. Calculations of the multi-group opacities were carried out using the Uppsala package \citep{Gustafsson08}. The output simulations describe a physical volume significantly larger than the size of granules enclosed in a grid with a typical resolution of 140 x 140 x 150 points. A detailed description of the simulation code, model setup, and calculations   can be found in \citet{Ludwig09}, \citet{Tremblay13}, and \citet{Ludwig21} along with descriptions of the limitations
of the models.

\begin{figure}[]
        \centering
        \resizebox{\columnwidth}{!}{\includegraphics[trim=20 0 40 25, clip]{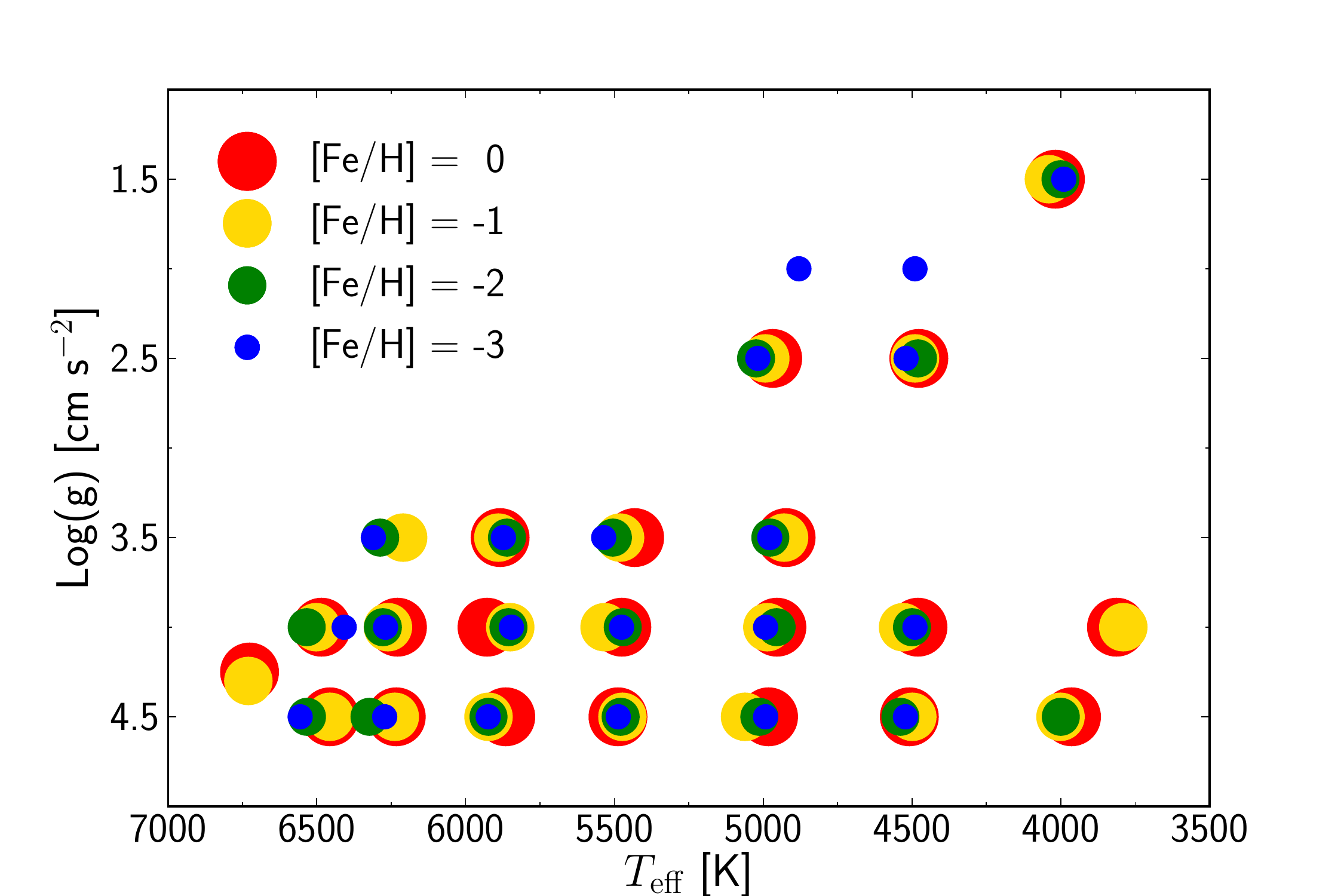}}
        \caption{Stellar parameters of the 84 three-dimensional models of atmospheres that make up the CIFIST grid.}
        \label{grid}
\end{figure}

The final grid of 3D models comprises 84 simulations with different combinations of surface gravity, chemical composition, and entropy flux at the bottom of the atmosphere. Figure \ref{grid} shows the coverage in stellar parameters of the grid. Effective temperature was determined from the temporally and spatially averaged emergent bolometric flux. For each 3D model, a 1D hydrostatic model of atmosphere was also derived by linear interpolation of the ODFNEW model grid by \citet{Castelli04}, which was also based on the solar abundances of \citet{Grevesse98}.

After completion of a model run, a subset of about 20 uncorrelated snapshots is selected for spectral synthesis purposes. The selection  is based on the requirement that the statistics of the subsample should  closely  resemble  the  statistics  of  the whole run. In particular, the statistics of the fluctuations in  velocity  and  emergent  flux  should be preserved. The total time-coverage spans at least ten times the typical lifetimes of convective cells on each modeled star.

\section{Spectral synthesis}

The calculation of stellar model atmospheres and spectra can be a time-consuming process. This is usually true for 3D models, but can also apply to  1D models when needed in large numbers,  as in the most recent grids for Kurucz \citep{Meszaros12}, MARCS \citep{Gustafsson08}, and Phoenix \citep{Husser} models, which can include up to hundreds of thousands of models. A widely extended technique to address this problem is to interpolate the main thermodynamical variables along the height of the atmosphere model. However, \citet{Meszaros13} showed that high-order interpolation of continuum-normalized fluxes leads to smaller errors than interpolation of stellar  atmospheric  structures. We expect the same result will hold for 3D models, and therefore we have decided to interpolate spectra instead of structures. This strategy is also much more efficient in the 3D case, given that 3D model atmospheres typically have between $10^4$ and $10^6$ more data points than their 1D counterparts.

For each 3D model atmosphere, we computed a synthetic spectrum making use of the radiative transfer code ASS$\epsilon$T, Advanced Spectrum Synthesis 3D Tool \citep{Koesterke09, Koesterke08}. The radiative transfer calculations on snapshots of 3D model atmospheres are a highly complex task compared to their 1D counterpart. This code is able to handle arbitrary line blends, frequency-dependent continuum opacities, and scattering. These calculations are described in a forthcoming paper \citep{Ludwig21}, but we provide a summary below for completeness.

The output spectra cover a wide range of wavelengths from 3000 to
30000 $\AA$.  For any given model, and for each frequency in the spectrum, a set of $\sim$2x10$^7$ points
in the temperature-density plane is required to provide detailed opacities for
every grid point in all the snapshots. This is dictated by the combination of the range covered by a given model in the temperature--density plane, and the required accuracy in the interpolated opacities. By assuming local thermodynamical equilibrium (LTE), this grid can be
built from the interpolation of a much reduced and pre-computed data set, with
steps of 250~K and 0.25~dex, using cubic B\'ezier interpolation. The adopted
solar abundances are very similar to the ones used in the 3D
models: as stated in the previous section, the simulations use the \citet{Grevesse98} except for CNO, for which the values from \citet{Asplund05} are adopted, and the spectral synthesis calculations embrace all the abundances from \citet{Asplund05}. As the main differences between Grevesse et al. (1998) and Asplund et al. (2005) are precisely in CNO, the changes are negligible for most purposes. Continuous absorption from H, H$^{-}$, and all relevant metals are taken into account, and line absorption is included in detail from the atomic and molecular files computed by Kurucz, with some upgrades, as described in detail in Allende Prieto et al. (2018). The opacities used in the hydrodynamical simulations are fairly consistent  with those adopted for the synthesis for temperatures above 3000 K, which are of concern for this work. The effect of these differences on model construction can be estimated from the work by \citet{Meszaros12}, and is considered minor.

Two sets of opacities were derived, one with spectral lines and high spectral resolution (2.7~m$\AA$ at wavelength 3000~$\AA$, adjusted with wavelength to keep sampling in velocity space constant), and the other one without lines and low resolution (1$\AA$ at 3000 $\AA$) in order to be used for the continuum spectra. For each individual snapshot, a spectrum is derived. First, the low-frequency pre-computed grid of opacities is interpolated to all grid points of the snapshot. A background radiation field is then calculated for this grid assuming LTE and including Thomson and Rayleigh scattering by atomic hydrogen. This radiation field is later used for the scattering term in the subsequent synthesis calculation. The line lists employed in the full-opacity calculations are the same as those used by \citet{Allende18}, and are available online\footnote{https://cloud.iac.es/index.php/s/pqJ9cwtJ9GSBraz}.

The emergent flux is finally calculated using the opacities that include
spectral lines and the previously computed mean background radiation field,
which is interpolated for all frequencies. The emergent flux is integrated
from the top layer down to optical depths larger than 20 for 21 different
angles: 4 equidistant azimuthal angles at an inclination $\mu = \cos \theta = 0.0886$, 8 at $\mu=0.4095$ and 0.7877, plus the vertical direction. Frequency shifts due to the velocity field are applied to the
opacities and source functions. The final emergent flux is an average from all the snapshots selected from the simulation.

In order to make a direct comparison between 1D and 3D spectra, we also derived 1D counterparts for the 3D spectra using a 1D version of the same radiative transfer code, ASS$\epsilon$T. The same chemical composition, opacities, and radiative transfer calculations are used, and a microturbulence of 1~km~s$^{-1}$ is adopted for all grid spectra. 

\section{Description of the interpolation tool}

Given the irregular nature of our grid of 3D models (see Fig.~\ref{grid}), we choose to use radial basis functions as the interpolation  method. A detailed description of these functions can be found in subsection \ref{rbfdescript}. 
The following two subsections are devoted to thoroughly describing the workflow and output of the code. As the interpolation is a computationally expensive process, and a dominant factor in the computations described, the calculation is divided into two steps. The interpolation weights or expansion coefficients are pre-computed and provided in tables together with the code, and the final step is done by the code, using the desired stellar parameters as an input.

The code and expansion coefficient tables are distributed from the CDS\footnote{And at https://cloud.iac.es/index.php/s/LKYXfYXtbqAKNfx}. The 3D spectra used in the code are preliminary, and not the final ones to be published in the 3D library \citep{Ludwig21}.

\subsection{Interpolation method}
\label{rbfdescript}

Every continuous function can be represented as a linear combination of functions:
 \begin{eqnarray}
f(x) = \sum_{j=1}^{m} w_{j}h_{j}(x),
\end{eqnarray}
where  $w_{j}$ are the expansion coefficients and $h_{j}(x)$ the blending or basis functions. Given a set of $x_{i}$ data points, their corresponding data values $s_{i}=f(x_{i})$, and a choice of basis functions, one can use Eq. (1) to write a linear system of equations. 

For multi-dimensional data, the above linear system may become singular. This problem can be addressed by defining a linear function whose parameters are the distances to the known data points \citep{Broomhead88}. Thus, the basis functions are radially symmetric around those data points. This approach is known as the radial basis functions (RBF) method, and the interpolation condition takes the form:
\begin{eqnarray}
f(x_{i}) = \sum_{j=1}^{m} w_{j}\phi(\vert\vert x_{i} - y_{j} \vert\vert) = s_{i},
\end{eqnarray}
where $x_{i}$ is a multi-dimensional data point, $s_{i}$ is the corresponding data value, $\phi$ is the RBF, and the norm is usually defined as the Euclidean distance between the data point and the centers of the RBFs, $y_{j}$, which are taken to be the known data points. We note that the function is constrained to go through the known data points. The flexibility of $f$, and its ability to fit many different functions, derives only from the freedom to choose different values for the weights, which are the only free parameters.

There are several types of RBFs that guarantee a unique solution for the system of equations \citep{Micchelli86}. One of them is the thin-plate splines (hereafter TPSs) 
  \citep{Duchon76}:
\begin{eqnarray}
\phi(r) = r^{2}\log(r),
\end{eqnarray}
where $r$ is the Euclidean distance: $r = \vert\vert  x_{i} - y_{j}
\vert\vert$ . TPSs have the advantage of producing smooth functions, as spectra are expected to be, with no free parameters to be tuned  manually. The final matrix representation of the interpolation condition is:

$$\begin{bmatrix}
  r^{2}_{1,1}\log(r_{1,1}) & r^{2}_{1,2}\log(r_{1,2}) & \cdots & r^{2}_{1,j}\log(r_{1,j}) \\
  r^{2}_{2,1}\log(r_{2,1}) & r^{2}_{2,2}\log(r_{2,2}) & \cdots & r^{2}_{2,j}\log(r_{2,j}) \\
  \vdots  & \vdots  & \ddots & \vdots  \\
  r^{2}_{i,1}\log(r_{i,1}) & r^{2}_{i,2}\log(r_{i,2}) & \cdots & r^{2}_{i,j}\log(r_{i,j}) 
 \end{bmatrix}
 \begingroup
\renewcommand*{\arraystretch}{1.15}
 \begin{bmatrix}
 w_{1} \\
 w_{2} \\
 \vdots \\
 w_{j}
  \end{bmatrix}
  =
   \begin{bmatrix}
 s_{1} \\
 s_{2} \\
 \vdots \\
 s_{i}
  \end{bmatrix}.
  \endgroup$$
  
The expansion coefficients $[w_{j}]$ are determined by solving this linear system of equations for the set of known data points.

\begin{figure*}[]
        \centering
        \resizebox{1\columnwidth}{!}{\includegraphics[]{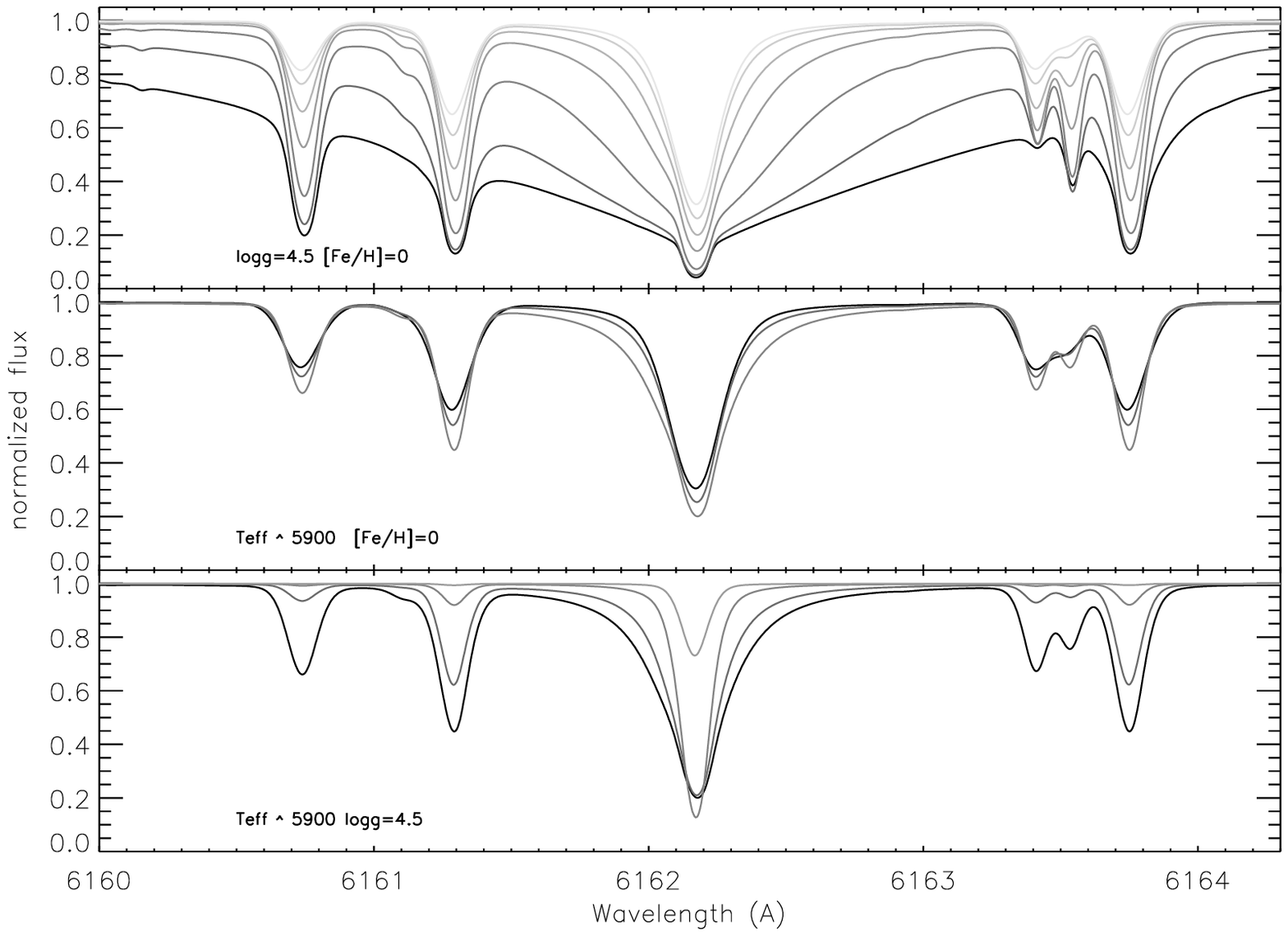}}
        \resizebox{1\columnwidth}{!}{\includegraphics[]{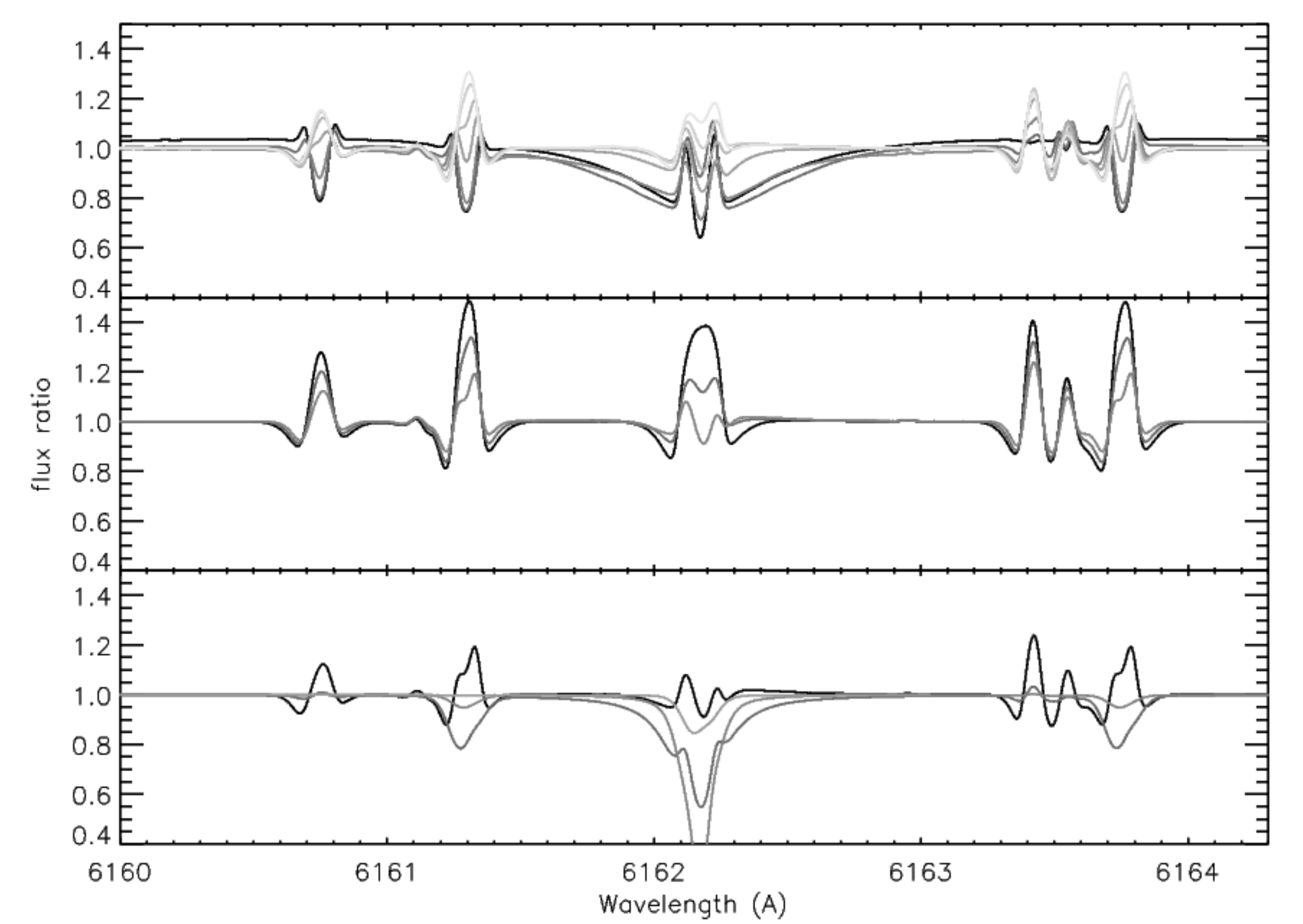}}
        \caption{Variations of the 3D flux (left-hand panels) and the ratio of the 3D and 1D model fluxes in the vicinity of the Ca I 6162.18 \AA\ line as a function of $T_{\rm eff}$ (top panels; $T_{\rm eff}$= 4001, 4499, 5061, 5473, 5923, 6238, and 6456 K), $\log g$ (middle; $\log g =$ 3.5, 4.0 and 4.5) and [Fe/H] (lower panels; [Fe/H]= -1, -2 and $T_{\rm eff}=5923$, -2 and $T_{\rm eff}=6287$ K, and -3).}
        \label{flux_and_fluxratio}
\end{figure*}

\subsection{Code workflow}
\label{sec:workflow}

We have implemented RBF interpolation in an IDL code which uses LU decomposition to solve the previous linear system. 
The code offers two different options to perform the interpolation: either interpolating the 3D normalized spectra, or the ratio between the 3D normalized spectra and their 1D counterparts (hereafter referred to as {\it straight} and {\it ratio} options of interpolation, respectively). Although stellar spectra show relatively smooth variations as a function of the stellar parameters, the ratio between these two types of model better captures the line asymmetries present in the 3D fluxes, as illustrated in Figure \ref{flux_and_fluxratio}, and this is the option we recommend. The code workflow described in this subsection corresponds to the {\it ratio} option of interpolation. The same procedure is applied to the {\it straight} interpolation option, but in this case the result of the {\it ratio interpolation} is finally multiplied by the interpolation in the internally stored 1D normalized spectra.

The spectra are provided at high resolution: between 3000 and 30 000~\AA ~the
number of wavelength points is on the order of $3.5\times10^6$, and each
wavelength is interpolated independently. The sampling is sufficient to resolve the velocity field, and it is equally spaced in frequency and velocity space. The same wavelengths are adopted for all the models. Due to limitations in computation time, and in order to provide an interpolated spectrum on the fly for practical use in research using a workstation or even a laptop computer, we divided the process in two steps. Therefore, some quantities have been pre-computed and stored.

As we describe in the previous subsection, to perform RBF interpolation, we have to build a matrix of weights (or expansion coefficients), and then solve a linear system of equations for a certain coordinates. The expansion coefficients have been
pre-computed. As the interpolation is performed independently for each wavelength, we have as many vectors of weights as wavelength points.

Our initial inputs are the 3D and 1D fluxes from the grid of spectra, and their corresponding continuums (see section \ref{sec:3dcont} for details of the 3D continuum corrections). We first selected a wavelength reference array, in this case the 3D spectrum with the highest resolution. This is necessary because the algorithm that optimally chooses frequencies for the 3D case produces slightly different results for each model and snapshot. All fluxes, 3D and 1D, were normalized and interpolated using cubic splines to this wavelength reference array. In addition, 1D spectra were convolved with a Gaussian profile to a macro-turbulence of 0.21 kms$^{-1}$ (0.5  kms$^{-1}$ FWHM)(see section \ref{sec:1dbroad} for details on the broadening of 1D spectra). Finally, the ratios between 3D and 1D were stored, and are used as the known data values, $s_{i}$.

In order to obtain the weights array, $w$, for each wavelength, we first built a matrix $M$ with the $r^{2}\log(r)$ elements, where $r$ represents the Euclidean distances between points in the grid. To do so, the stellar parameters should be first scaled due to the significantly different scales ($\log g$ has steps in the grid that are $10^3$ larger than $T_{\rm eff}$), and were normalized between 0 and 1. Subsequently, we solved the linear system $Mw=S$, where $S$ contains the ratios of normalized fluxes 3D/1D of all models,  $s_{i}$, at a certain wavelength. The final vector of weights contains 84 elements, one per model. All the vectors for all wavelengths are stored in a FITS table that can be found in the software package.

The IDL code that is available for the user is devoted to the second part of
the interpolation. The user provides a wavelength range  as input, together with a set of
stellar parameters: $T_{\rm eff}$, $\log g$, and [Fe/H], and the wavelength and flux arrays of a corresponding 1D spectrum, along with its resolution. In the case where no 1D spectrum is provided, {\it straight} interpolation will be performed. For each wavelength, a matrix with the $r^{2}\log(r)$ elements is built, where $r$ are the distances from the grid points to the set of stellar parameters provided. Afterwards, the linear system with the stored weights is solved using the matrix built earlier.

The evaluation of the linear system provides the interpolated 3D/1D flux ratio. As both the 3D and 1D fluxes were obtained with the same code and opacities, we expect these corrections to be of general application, even to different families of similar 1D models. Obviously, the closer the line opacities used are to those adopted in the computation of the ratios, the more accurate the results will be. In order to recover the 3D spectrum, we should multiply this quotient by a 1D spectrum; the code can take two arrays of wavelength and 1D normalized flux as input to do so. The input 1D spectrum should have the same micro- and macroturbulence as the ones in our grid: $\xi=$1 kms$^{-1}$ and $\upsilon_{mac}$=0.5 kms$^{-1}$ (FWHM). The code then reduces the resolution of the ratio by convolution with a Gaussian kernel to equate it to the input 1D spectrum, and then multiplies the two. Otherwise, a 1D Kurucz model from the CIFIST grid is interpolated for this purpose using the same RBF interpolation. It is important to note that, as the  computation of 1D spectra is less expensive, it would be more precise and recommended to compute the 1D model specifically for the required stellar parameters, and not to interpolate it. It is important to stress that the wavelength interval for interpolation is, within the range available in the model grid (3000--30,000 \AA), entirely the choice of the user. As the RBF interpolation is performed independently for every wavelength, the size of the wavelength window has no impact on the accuracy of the interpolation at any particular wavelength in that window.

In summary, the code provided can be used to predict 3D corrections for any given input 1D spectrum. The code will interpolate the precomputed 3D/1D flux corrections, smooth them to the appropriate resolution of the input 1D spectrum, and multiply them by the input 1D spectrum to return an approximate predicted 3D  spectrum. The code can also work in a simplified mode, in which the output 3D spectrum will be returned when no 1D input spectrum is provided. We believe the methodology of storing and interpolating 3D/1D flux corrections is of general application for the practical determination of chemical abundances and stellar parameters taking into account hydrodynamical effects from the analysis of stellar spectra. In this sense, the code provided with this work can be considered a prototype for future applications.

\begin{figure}[]
        \centering
        \resizebox{0.8\columnwidth}{!}{\includegraphics[]{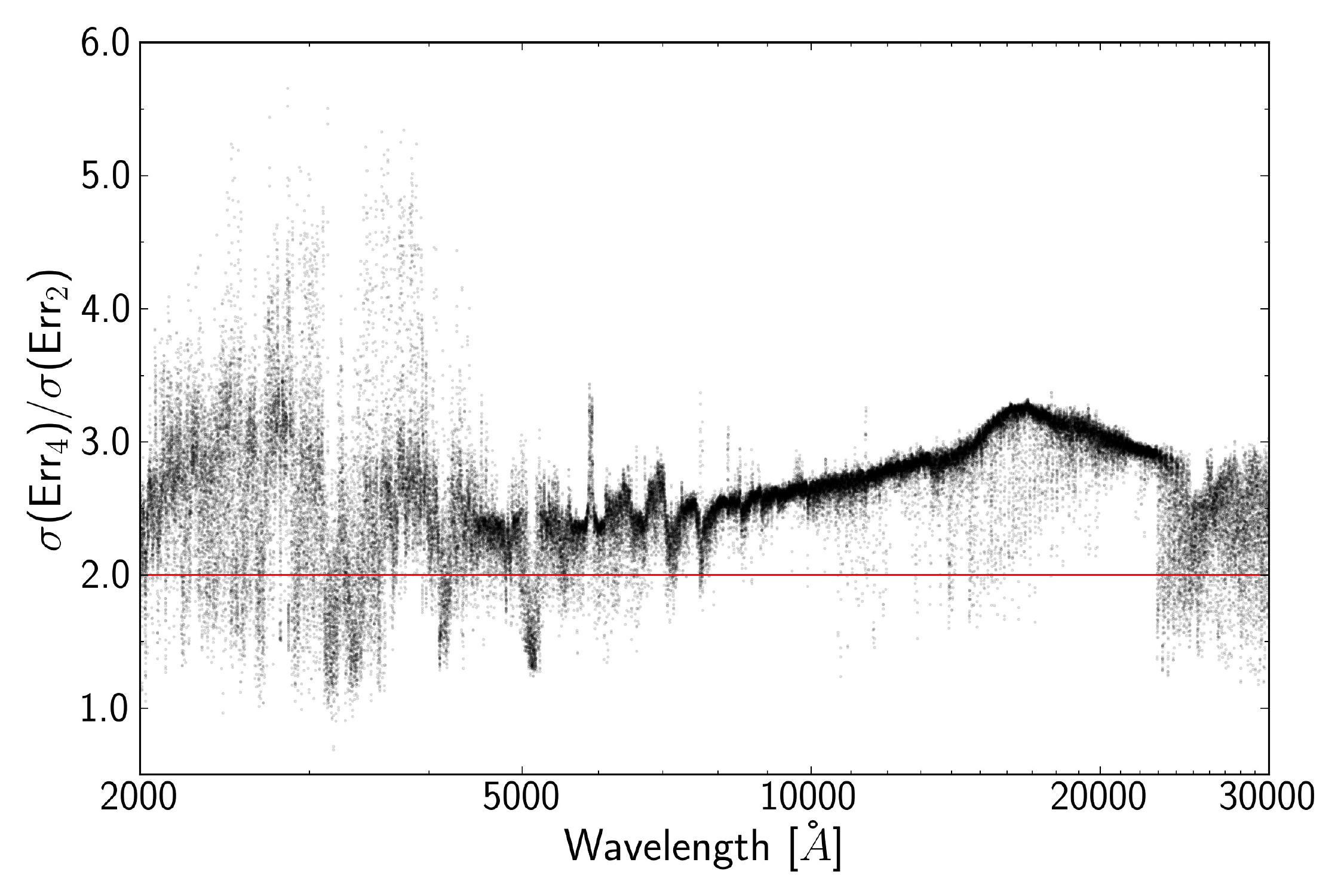}}
        \caption{Quotient between the standard deviation (STD) of errors derived from the interpolation using the quarter subgrid as the model, and the STD of the errors using the half subgrid as model, for all wavelengths. A horizontal line is drawn at 2.0.}
        \label{redfact}
\end{figure}

\section{Interpolation errors}
\label{sec:errors}

In order to provide an estimation of the performance of the interpolation model, we carried out a leave-one-out cross-validation. We systematically left out from the dataset one 3D/1D quotient, and used the remaining  n-1 elements in the grid to derive the interpolated 3D/1D ratio corresponding to the stellar parameters of the model that was left out. Afterwards, we computed the differences between the original data and the interpolated ratios. 

To finally obtain the differences between the  spectra of the 3D models and their interpolated counterparts, we multiplied the quotients by the corresponding 1D spectrum. Those differences were adopted as the upper limits of the errors for the interpolated spectra at the points of the grid. This process was repeated for all models and all wavelengths. The errors between grid points are evaluated by interpolating the errors derived from the cross-validation exercise  using the same method. For {\it straight} interpolation, the same procedure is followed, but here the interpolated ratio is multiplied by a spectrum interpolated in the grid of 1D spectra.

The RBF interpolation uses spline functions, which means that the interpolated spectra at the grid points will always correspond exactly to the value of the point. The leave-one-out cross-validation generally gives a pessimistic estimate of the performance of a model, due to the fact that the model is trained with less grid points. In our particular case, the distance to the nearest grid points is typically doubled when we suppress a data point. To constrain and provide more realistic errors, we performed the following test to estimate how errors scale with  the step size of a grid.

For statistical significance, we used an extensive 1D grid of 700 spectra  spanning approximately  the same parameter range as the 3D grid. We created two subgrids, one taking half the points from the original grid, and a second  again taking half of the points from the first (a quarter from the original grid). Grid points are equidistant in $T_{\rm eff}$, $\log g$, and [Fe/H]. The distance between points of the one-quarter grid is doubled compared to the one-half grid. For all the wavelengths from 2000 to 30000~\AA, we interpolated the spectra at 175 points from the original grid using both the one-half and the one-quarter grids as input. Finally, for each frequency, we calculated the standard deviation of the errors, that is, the difference between the interpolated and the data flux. The quotient between the standard deviation (STD) for the one-quarter grid and that for the one-half grid is over 2.0 for the vast majority of frequencies (see Fig. \ref{redfact}).

We stress that this experiment is only targeting the derivation of the scaling factor in the uncertainties as a function of the grid step size. This factor needs to be applied to the independent leave-one-out experiment described at the beginning of this section. As the scaling factor is typically in the range between 2 and 3, we conservatively divided the errors from the cross-validation by a factor of 2 to account for the doubling of the maximum distance between grid points during the interpolation process. There is no reason to expect a different scaling behavior for 3D and 1D models.

Despite the fact that we divide the error estimates from the leave-one-out tests by a factor of 2, it is important to note that at the edges of the interpolation grid these are still overestimated because of the nature of the leave-one-out cross-validation process. Once the grid point on the edge is included, and the interpolation is performed within the grid, the accuracy of the interpolation should improve. As we see below in practical applications, the uncertainty in the interpolations is generally under 2 \%, although there are particular regions of the parameter space where the accuracy degrades somewhat.

\begin{figure*}[]
        \centering
        {\includegraphics[trim=25 470 25 10, clip]{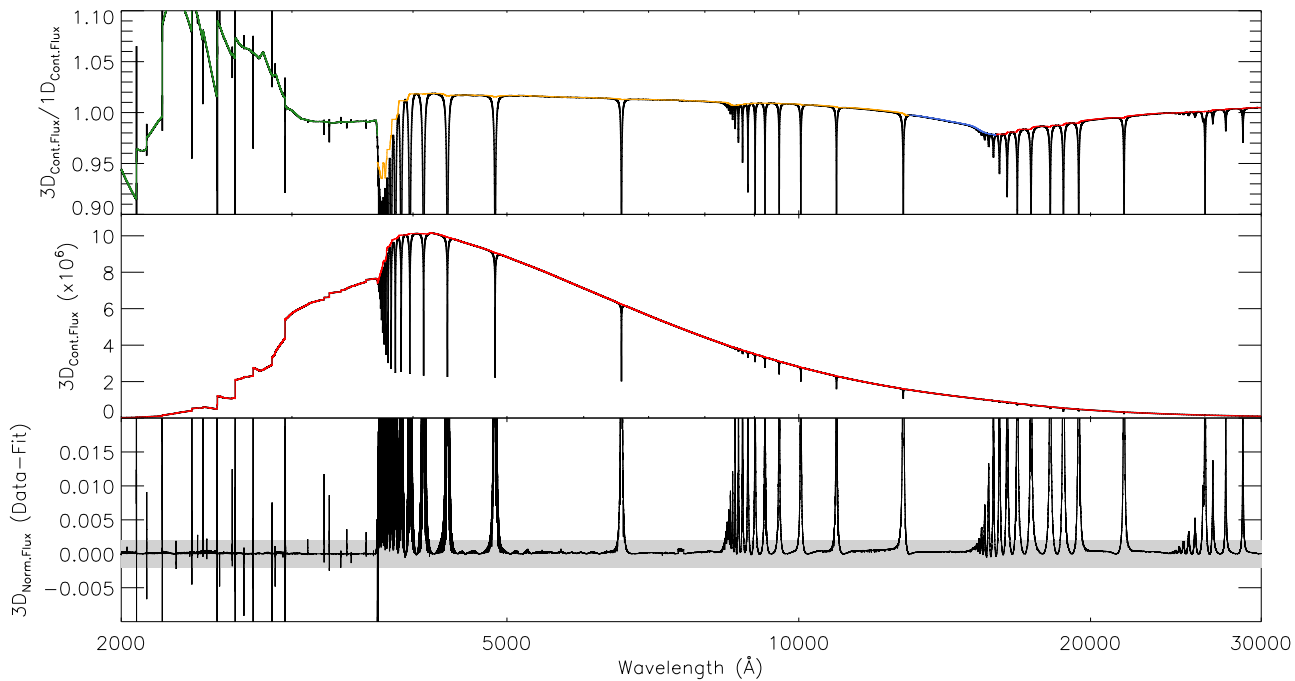}}
\caption{3D continuum correction of the model with stellar parameters $T_{\rm eff}$=5865K, $\log g$=4.5, and [Fe/H]=0.0 dex. \textit{Upper panel:} Fit to the upper envelope of the quotient between the 3D and 1D continuums. The different wavelength ranges used for the fit are plotted in different colors. \textit{Middle panel:}  Original 3D continuum, shown in black, while the new continuum fitted is depicted in red. \textit{Bottom panel:} Difference between the 3D spectrum normalized with the original continuum, and the 3D spectrum normalized with the new fitted continuum. The shaded area corresponds to the $\pm$0.2\% of the normalized flux.  }
\label{continuum}
\end{figure*}

\section{Limitations of the tool}

There are two main issues concerning the treatment of the spectra preceding the interpolation. These should be noted because they can diminish the precision of the final interpolated spectra. Nevertheless, as we explain in this section, the introduced errors are actually negligible for most wavelengths.

\subsection{3D continuum correction}
\label{sec:3dcont}

As noted above, the interpolation tool takes as input  a
  normalized flux spectrum. However, continuum spectra for the 3D models were computed
including the opacity from hydrogen lines, while the 1D continua were
calculated without them\footnote{The differences between 1D and 3D continua
  regarding the hydrogen lines stem from the design of the software package
  used to compute the 3D opacities}. While including hydrogen lines as continuum is useful to 
  analyze lines sitting in their wings, in most cases these lines are not used and, on 
  the other hand, the wings of hydrogen lines can provide valuable information.
  In order to provide useful theoretical
spectra to be compared with real stars, hydrogen lines must be present in the final
interpolated spectra, and therefore they should be removed from the continuum
before normalization. 

We clean hydrogen lines from the continuum by first dividing the 3D continuum by its 1D  counterpart to make the hydrogen lines stand out and reduce the error in this procedure. Afterwards, we apply a moving percentile filter to this quotient in order to fit its upper envelope. Specifically, for each wavelength we select an interval around it of a certain width, and calculate the distribution of the corresponding ratios of 3D and 1D continuum fluxes. We then select the desired percentile of the distribution and assign it to the central wavelength.

The values of the width and percentile for the filter were customized for four different wavelength intervals and for each spectra individually. The wavelength intervals are slightly different from spectrum to spectrum, but cuts were approximately made at 3675, 13000, and 16000~\AA. The width of the interval was around 2000 points for the UV and from 2·10$^{4}$ to  7·10$^{4}$ points for the remaining wavelengths. Finally, the percentile was chosen to be 70$\%$ in the UV, and around 90$\%$ at longer wavelengths.

As an example, Figure~\ref{continuum} shows the fit to the upper envelope of the continuum ratios, as well as the different wavelength intervals created to fit the quotient, for a simulation with solar-like parameters. The final 3D continuum without hydrogen lines is shown superimposed over the original one in the middle panel. To finally obtain the errors associated with fitting the continuum that we need to add to areas without a hydrogen line, we plotted the difference between the normalized fluxes when using the original continuum and the corrected one. The lower panel of Figure~\ref{continuum} displays those differences, where hydrogen lines that were missing in the original normalized 3D spectra now appear in the new normalized spectra. As can be seen in the figure, outside the areas influenced by the wings of hydrogen lines, the differences from the original spectra are smaller than 0.2\%.

Because of the lack of smoothness in the continuum fit to the Balmer
discontinuity, the spectra around these wavelengths may not be reliable, and should be used with caution.

\subsection{Broadening of 1D spectra }
\label{sec:1dbroad}

\begin{figure*}[]
        \centering
        \resizebox{2\columnwidth}{!}{\includegraphics[trim=00 520 0 00 00, clip]{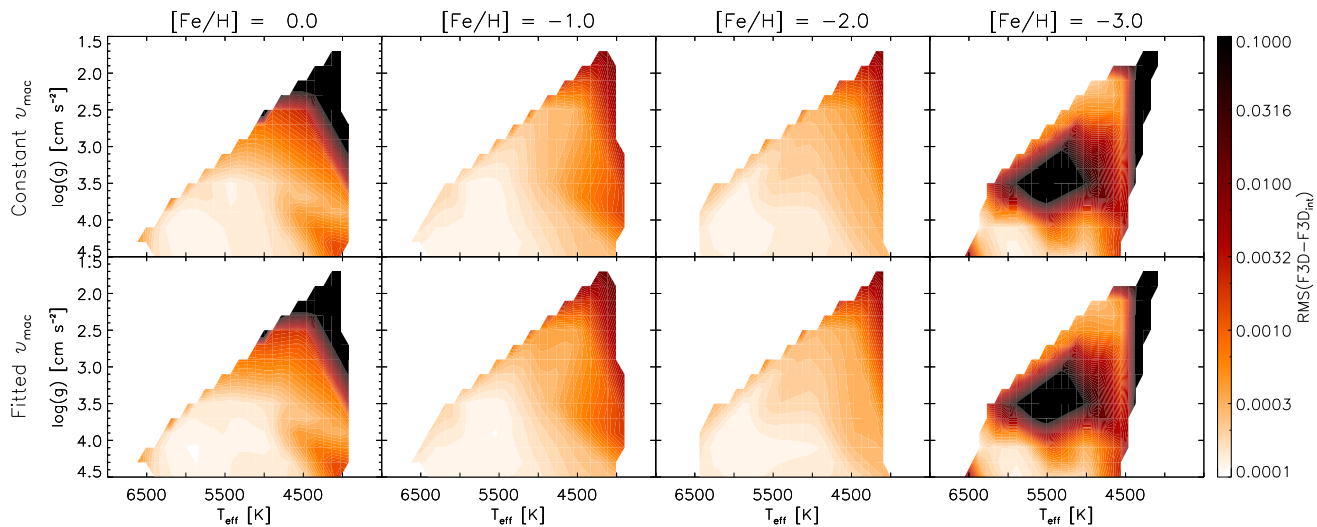}}
\caption{RMS of the differences between the 3D flux and the interpolated 3D flux when the corresponding point is subtracted from the grid. The upper row corresponds to interpolations where all 1D spectra were broadened with the same macroturbulence, while for the lower panel, the macroturbulence was individually fitted for each spectra. A logarithmic color scale is used for plotting purposes.}
\label{macrorms}
\end{figure*}

The second main issue concerns 1D spectra. Theoretical classical models are computed with a certain microturbulence, and should afterwards be convolved with the corresponding macroturbulence. Doing so, we mimic the broadening of spectral lines associated with velocity fields missing from 1D models because of their hydrostatic nature. Although there are experimental relationships between macroturbulence and effective temperature \citep[e.g.,][]{Valenti05}, no theoretical determinations are currently available. The absence of turbulent or velocity broadening in 1D spectra could therefore enhance the differences between 1D and 3D model fluxes, making the interpolation more difficult. Hence, we have to select certain micro- and macroturbulence to broaden the 1D spectra prior to interpolation.

In order to select appropriate values for the macroturbulence, we performed the same test for two different sets of this parameter. Our first approach was to select the macroturbulence for each 1D spectrum that minimized the differences between this latter and its corresponding 3D spectrum. The second approach was to select the same macroturbulence for all models. This constant value should be enough to smooth the spectrum, removing small features that are never visible and not high enough to blur spectral lines. We choose  and $\upsilon_{mac}$=0.5~kms$^{-1}$ (FWHM), using a Gaussian profile for the convolution. This value may seem too small; a typical value for the Sun is $\upsilon_{mac}$=3.4 kms$^{-1}$ \citep{Allende01}. However, the grid covers a wide range in stellar parameters, and we have to choose a value that does not blur the spectral features in any of the stars included in the parameter space of the grid. For both cases $\xi$=1~kms$^{-1}$. 

With these two different sets of parameters ($\xi$, $\upsilon_{mac}$), we broadened each 1D spectrum to be later used in the interpolations. We selected for the test the wavelength range from 6156 to 6173~\AA,  populated by a strong Ca~I line and other features. Due to its relative complexity and range of line depths, this region challenges our interpolation tool. For each 3D model, we created a subgrid with all the 3D/1D quotients except the one corresponding to the selected model. We then used this subgrid to calculate the interpolated 3D/1D ratio corresponding to the stellar parameters of the selected model. The ratio was then multiplied by a 1D spectrum with the same micro and macroturbulence as the modeled one in order to recover the 3D interpolated spectrum. Finally, we derived the RMS between the interpolated spectrum and the original 3D spectrum that was  dropped from the grid. 
 
We found that the derived RMS was almost equal for all sets of micro
and macroturbulence values, which indicates that the selection of these
parameters has little impact on the interpolation results. As  the choice must be consistent with the  input 1D spectrum that is to be multiplied by the 3D/1D ratio in order to derive the 3D spectrum, one would naively expect the choice to be irrelevant, and this is exactly what this test demonstrates.
 
Figure ~\ref{macrorms} illustrates the results of this test. The first row of plots corresponds to interpolations performed with 1D spectra broadened with constant $\upsilon_{mac}$, while the second row corresponds to 1D spectra broadened with a fitted $\upsilon_{mac}$ for each spectrum. Despite small differences that can be seen in the areas corresponding to high $T_{\rm eff}$ and $\log g$, both set of plots are remarkably similar. Therefore, we adopted the simplest option: constant values for both the microturbulence and macroturbulence.

\section{Code output}
\label{sec:output}

The code will provide three arrays with the 3D interpolated and normalized
flux, associated wavelengths, and an array of interpolation errors. In addition, the code can produce a two-page document with several graphics intended to help the user to evaluate the 3D effects, the quality of the interpolation, and the improvement of using 3D interpolations against 1D theoretical spectra in the specified wavelength range. This information allows users to evaluate whether, despite the errors in the interpolation, the 3D interpolated spectra is a fair approximation of an exact 3D calculation and whether such calculation differs significantly from a 1D model. In particular, the plots produced (see section \ref{sec:apps} for examples and further details) are: (1) Mean(F3D/F1D): In order to check for possible 3D effects present in a given wavelength range, we plot contour maps of the mean normalized flux ratio 3D/1D averaged over all requested wavelengths, for each [Fe/H] value in the grid. (2) RMS(F3D-F1D): In order to compute the differences between 3D and 1D, we produce contour maps of the RMS of the difference between the 3D and 1D normalized fluxes, at the wavelength range requested and for each [Fe/H]. (3) RMS(F3D-F3D$_{int}$)/2: Contour maps of the RMS of the difference between the 3D normalized flux and its 3D interpolated counterpart (removing the corresponding point of the grid) in the wavelength range requested and for each [Fe/H]. A reduction factor of 2 is applied to this difference (see Section \ref{sec:errors}). (4) RMS(F3D-F1D)/RMS(F3D-F3D$_{int}$)/2: Contour maps of the ratio between plot 2 and 3. The aim of this plot is to show which is a better approximation to actual 3D calculations, the 1D spectrum
or the 3D interpolation, and quantify the improvement. If 1D fits better, contours appear in grayscale. It is not recommended to use the interpolation in this case.
(5) Mean(|F3D-F3D$_{int}$|/2): The errors of the interpolation are shown in these graphs, where contour maps are plotted for the mean absolute error, averaged over all requested wavelengths, and for each [Fe/H] value.
(6) Result of the interpolation: 1D spectrum (either provided or interpolated) and 3D interpolated spectrum with the corresponding errors.

It is important to take into account the fact that the described graphics provide
quantities averaged for the whole wavelength range. If the wavelength range requested is too wide, continuum regions will dominate the result, and any 3D effect will be diluted in the contour plots. In addition, wavelength-averaged errors will also be smaller than those for selected parts of lines, for example when only the core or the wings of a spectral line are considered. Finally, we note that the 1D spectrum used for the contour plots is not the one provided by the user, but the 1D from our grid computed for the same parameters as the 3D one.
The following section shows practical examples of application of the code, its input and output data, and the output graphics.

\section{Output examples}
\label{sec:apps}

We selected four spectral regions of interest to test the performance of the interpolation code. For each region we show the output diagnostic plots described in Section \ref{sec:output}. These examples also highlight some issues that should be taken into account for proper interpretation of the diagnostic plots.

It is important to note that the 1D spectra used for the comparisons have hardly been broadened, with an adopted macroturbulence of $\upsilon_{mac}$=0.5 kms$^{-1}$ (FWHM). Therefore, in regions of the $T_{\rm eff}$-$\log g$ plane where macroturbulence is much greater (i.e., for high temperatures and low gravities), 3D effects may appear artificially enhanced. A more adequate comparison would involve broadening the 1D spectra with appropriate values of the micro- and macroturbulence for each $T_{\rm eff}$ and $\log g$.

\begin{figure*}[]
   \centering
   \resizebox{1.5\columnwidth}{!}{\includegraphics[trim=0 35 0 20, clip]{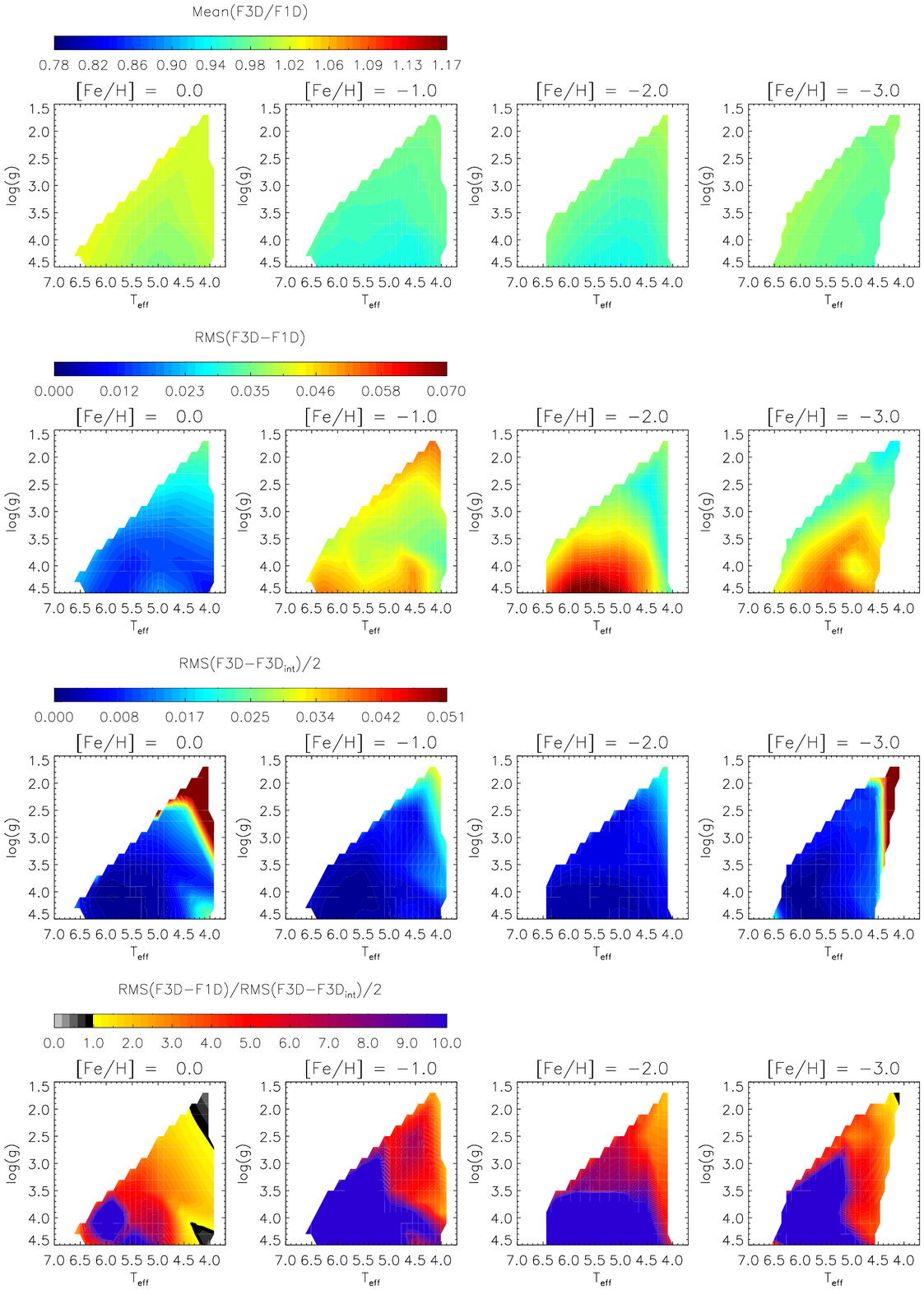}}
   \resizebox{1.5\columnwidth}{!}{\includegraphics[trim=0 410 0 40, clip, width=0.1\textwidth]{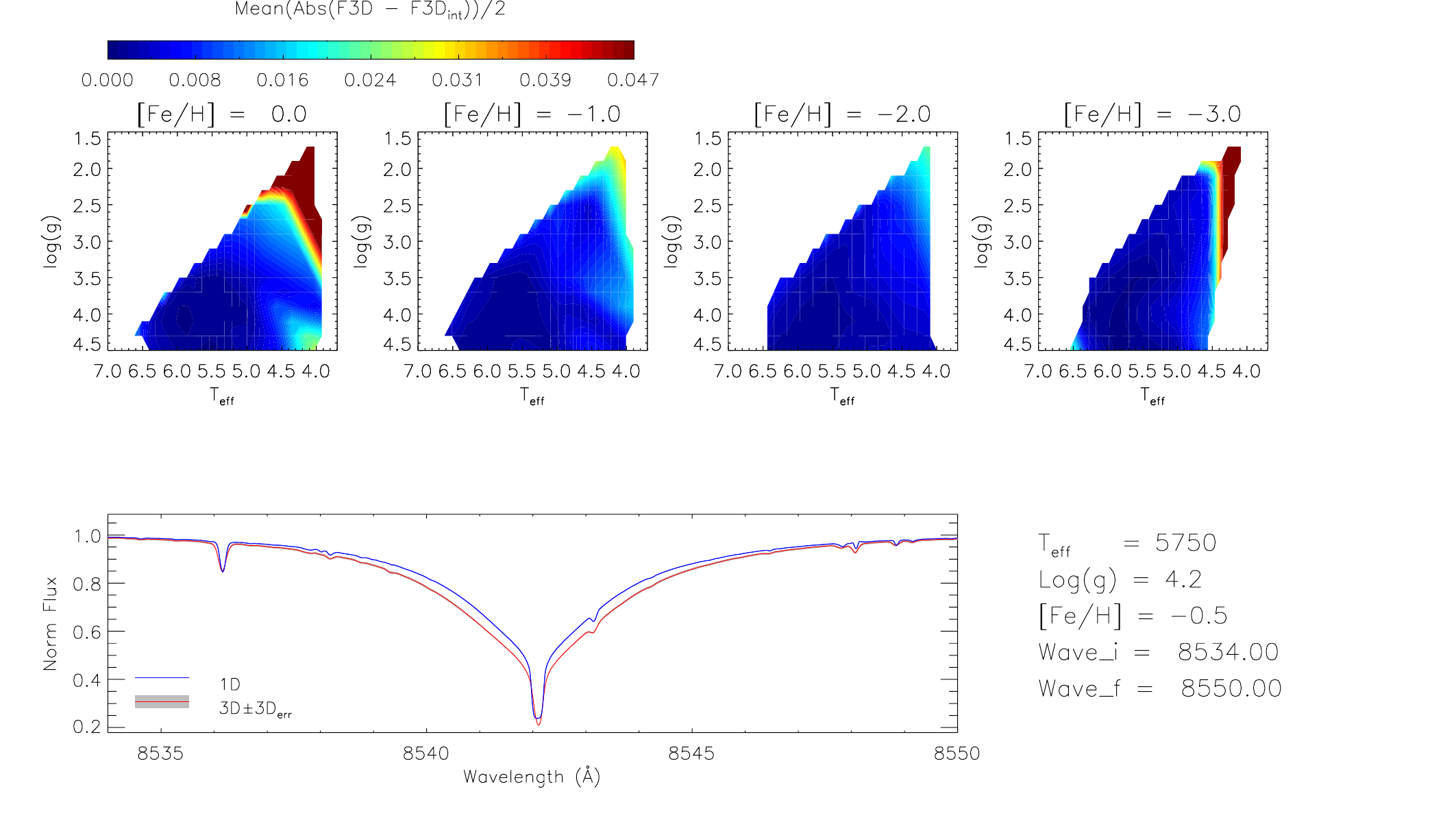}}
  \caption{Diagnostic plots for the Ca II spectral line at 8542.09~\AA. Effective temperature is expressed in $\times10^3$~K. We note that in the bottom-right panel, the uncertainties (gray shadow) are not visible due to their small size.}
   \label{fig:diagCa}
\end{figure*}

\begin{figure*}[]
   \centering
   \resizebox{1.5\columnwidth}{!}{\includegraphics[trim=0 35 0 20, clip]{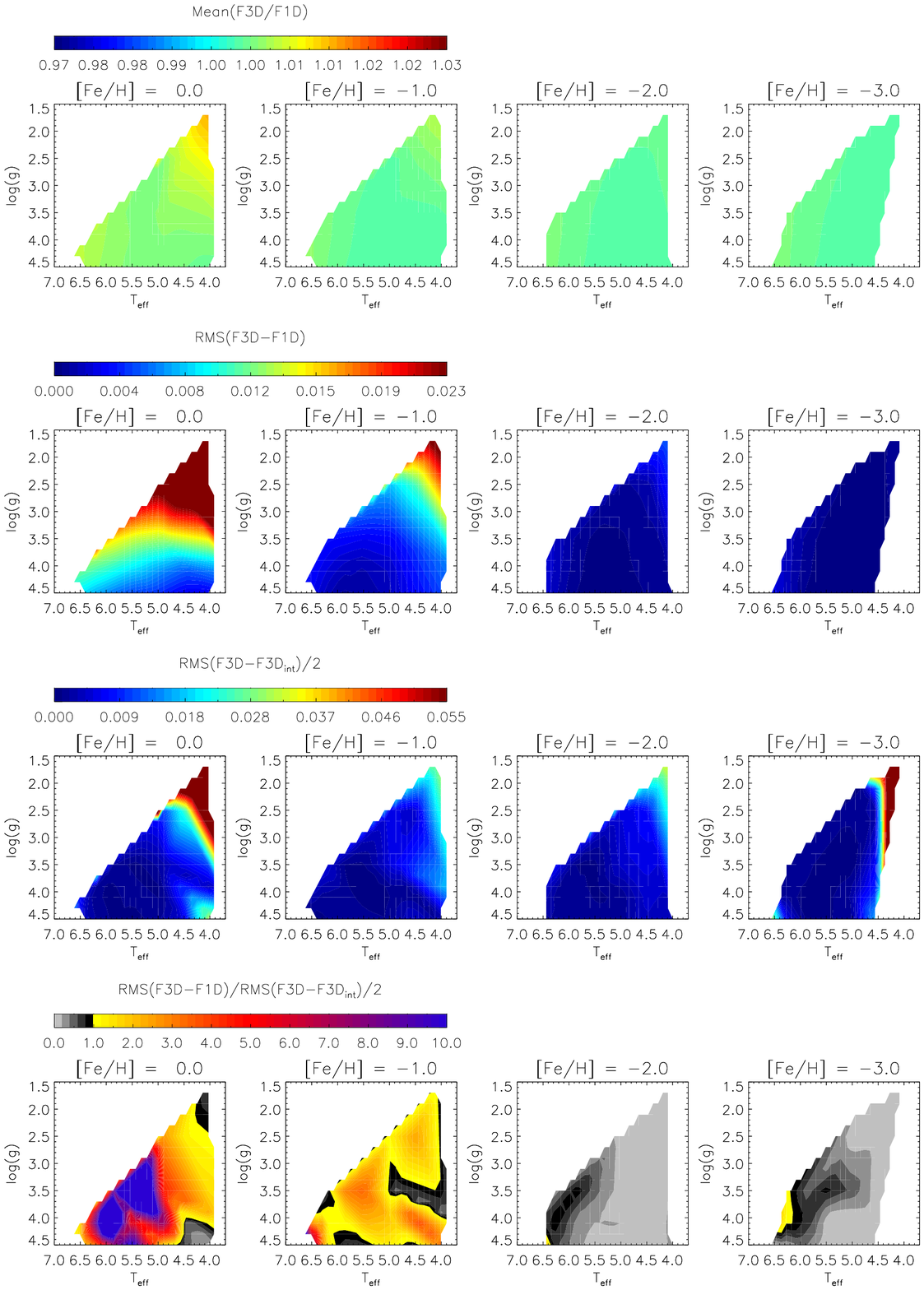}}
   \resizebox{1.5\columnwidth}{!}{\includegraphics[trim=0 410 0 40, clip]{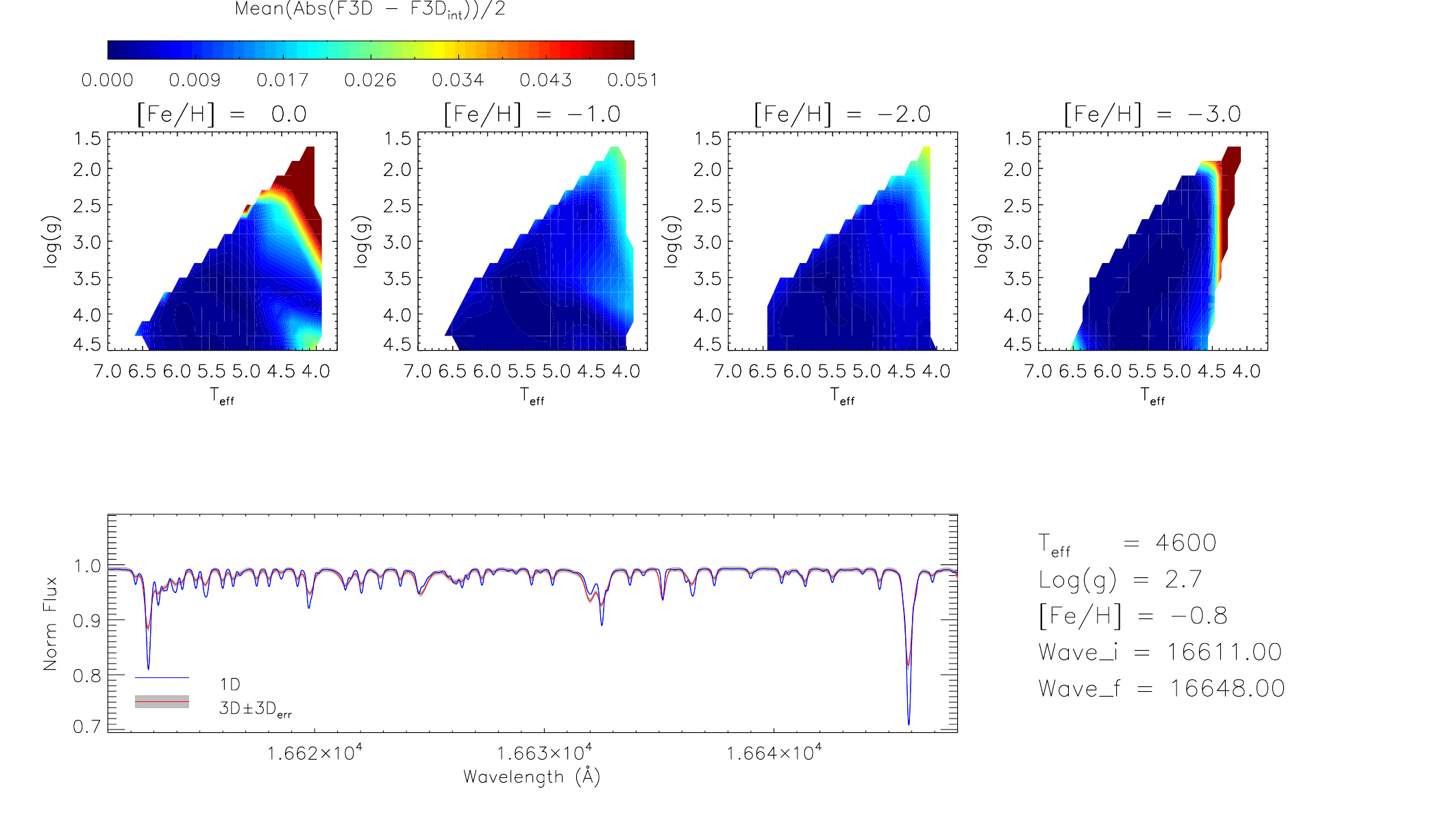}}
  \caption{Diagnostic plots for the CO molecular band around 16611-16640~\AA. Effective temperature is expressed in $\times10^3$~K.}
   \label{fig:diagCO}
\end{figure*}

The first two rows of contour plots in Figures \ref{fig:diagCa} to \ref{fig:diagHa} are shown in order to highlight 3D effects averaged over the selected wavelength range. The first one shows the mean ratio 3D/1D for the model spectra in the grid, and the second one depicts the root mean square (RMS) between the 3D and the 1D model spectra. We selected the RMS as a measure of the average differences in flux, regardless of their direction (positive or negative). We warn the user that minor systematic differences between the 3D and 1D continua may be present, which can increase the actual 3D-1D differences. These are real differences, some of which can survive the normalization process. They are mostly undetectable in the optical and near-infrared (NIR), but become prominent in the UV, or at longer wavelengths near the Paschen or the Brackett series. The RMS plots will be more sensitive than the 3D/1D plots to small discrepancies between models, which are typically what most users are interested in. In these figures, the range of the color bar in all cases covers $\pm2\sigma$ from the average. These two first types of plots are independent of the interpolation method used ({\it straight} or {\it ratio}), and simply provide a glance at possible 3D effects.

The third row of the contour plots contains the result of the leave-one-out cross-validation performed to find the errors of the interpolation. For each point of the grid, the 3D interpolated (3D$_{int}$) spectrum is computed using a grid that contains all spectra except the one we are interpolating. We then calculate the difference between the model and the interpolated spectra. Finally, in order to avoid an overestimation of the errors, errors are divided by the reduction factor 2 (see Section~\ref{sec:errors}). The contour plots show the RMS of this difference (F3D-F3D$_{int}$)/2, which will change depending on the interpolation method that is used. For these figures, the {\it ratio} interpolation option is always used.

The fourth row of panels in the plots depicts the ratios between quantities in
rows two and three. When the difference F3D-F1D is lower than
(F3D-F3D$_{int}$)/2, the 1D model spectrum is closer to the
3D model than the 3D$_{int}$. In that case, the interpolation is not improving
the model spectrum but the opposite, and is depicted in grayscale. There are
two main reasons why this can happen: either errors in the
interpolation are significant, and larger than the 3D-1D differences, or we
are close to the edges of the grid, where the interpolation errors are probably overestimated due to the fact that an extrapolation was performed to compute them. When the 3D$_{int}$ flux fits the 3D model  better, contours are depicted in color scale.  For these plots, the range of the color bar will always be between 0 and 10.

Finally, the last contour plots show the mean errors of the interpolation. These errors cannot be understood as a standard deviation of a statistical distribution of errors, but as an absolute measurement of the deviation of the interpolation from the 3D model. As for all other contour plots, it is an average over the wavelength range of interest.

To conclude with the series of diagnostic plots, the 3D interpolated spectrum is shown together with its errors for the stellar parameters specified by the user. The 1D spectrum is also plotted, that is, either the one provided by the user or the 1D spectrum interpolated from our grid, with $\upsilon_{mac}$=0.5 kms$^{-1}$ and $\xi$=1kms$^{-1}$, which is the case for the figures in this paper. In addition, the wavelength range and stellar parameters are shown.

\subsection{Infrared Ca triplet}

The first wavelength range selected for the tests comprises the strongest line from the Ca II  IR triplet, at 8542.09~$\AA$ (Fig. \ref{fig:diagCa}). This triplet is broadly used in stellar spectroscopy because of its strength, its detectability over a wide range of atmospheric parameters, and the fact that it lands in a region where CCD detectors are highly efficient. It is the focus of massive spectroscopic surveys such as RAVE \citep{2020AJ....160...83S} or the Radial Velocity Spectrograph of the  Gaia mission \citep{2018A&A...616A...6S,2019A&A...622A.205K}.

As we can see from the first rows of panels, despite the line becoming weaker, 3D effects become more significant as we move towards lower metallicities. At [Fe/H]$<$0, the spectral line has a larger equivalent width (EW) in the 3D case, because 3D/1D$<$1, and differences in flux can reach up to 6\% for dwarf stars at [Fe/H]=-2. Although 1D spectra are not broad enough, macroturbulence does not change the EW of the line, and therefore we can safely conclude that at subsolar metallicity 1D Ca abundances would be overestimated.

The performance of the interpolation is satisfactory for all stellar parameters given the reduced RMS of the 3D-3D$_{int}$, which is under 3\% for almost the whole grid. However, at the lower gravity end of the grid, the interpolation is not reliable. As mentioned above, this is probably due to the fact that it was necessary to do an extrapolation in these cases to derive the 3D$_{int}$ spectrum during the leave-one-out cross-validation, and therefore the difference 3D-3D$_{int}$ will appear overestimated around these low-gravity points of the grid.

The fourth line of contour plots shows that, in this case, the 3D interpolation is highly recommended, as it reproduces the real 3D spectrum four times better than the size of the 3D-1D differences. In addition, errors are below 0.12\%, with the exception of the low-gravity edges of the grid.

The last plot shows the 3D interpolated spectrum together with its errors, which are visible in the wings of the line as shaded areas around the spectrum, although they are nearly invisible in this case, hidden by the 3D curve. The 1D spectrum ($\upsilon_{mac}$=0.21 kms$^{-1}$ and $\xi$=1kms$^{-1}$) is also depicted for comparison. The 3D interpolated spectrum has wider wings, and clearly shows a redshift of the core. This is a limitation of 3D model atmospheres ---and is not related to the interpolation---, which show redshifted cores in strong lines with EW$>$100~\AA\  that are not observed in stars \citep{2009MmSAI..80..622A}.

\subsection{CO molecular band}

We select a CO molecular band in the IR between 1.6611 and 1.6648~$\mu$m. CO features in the $K$ and $H$ bands are frequently used to measure carbon abundances in late-type stars, and carbon isotopic ratios. This particular band is now of great interest because it is included in the spectral range of the APOGEE survey \citep{2017AJ....154...94M}, which offers a growing public data base with nearly half a million spectra \citep{2020AJ....160..120J,2021arXiv211202026A}.

The corresponding diagnostic plots are shown in Fig.~\ref{fig:diagCO}. Differences between 3D and 1D arise at high metallicities, and are particularly significant for giants. This may be partially caused by the low micro- and macroturbulence used to calculate 1D spectra. On the other hand, the 3D-1D differences seen in the contour plots at low metallicity are minor, because the molecular band is barely detected in this case.

The 3D interpolation is relatively reliable for all stellar parameters in light of the RMS(3D-3D$_{int}$) contour plot. Regarding the Ca II example, the low-gravity end of the grid tends to present  interpolations of poorer quality, with larger errors. This will probably be the case regardless of wavelength, because the extrapolation of these sharp ends of the grid during the leave-one-out cross-validation may be of very poor quality.

Despite the values of RMS(3D-3D$_{int}$) suggest the results are reliable, we find that  interpolation of 3D models is not recommended at [Fe/H]$\leq$-2, or  even at [Fe/H]=-1, for most of the $T_{\rm eff}$-$\log g$ pairs of parameters, given that the 3D-1D differences are very small.

Finally, the CO band is shown for a cool subgiant at [Fe/H]=-0.8. Errors, as already depicted in the contour plots, are around 1\% for this set of parameters. The 1D spectrum clearly lacks broadening, although a proper macroturbulence convolution by itself would not reconcile the differences with the 3D spectrum. The strong features at the edges of the wavelength range correspond to Fe I transitions.

\subsection{Lithium resonance line}

\begin{figure*}[]
   \centering
   \resizebox{1.5\columnwidth}{!}{\includegraphics[trim=0 35 0 20, clip]{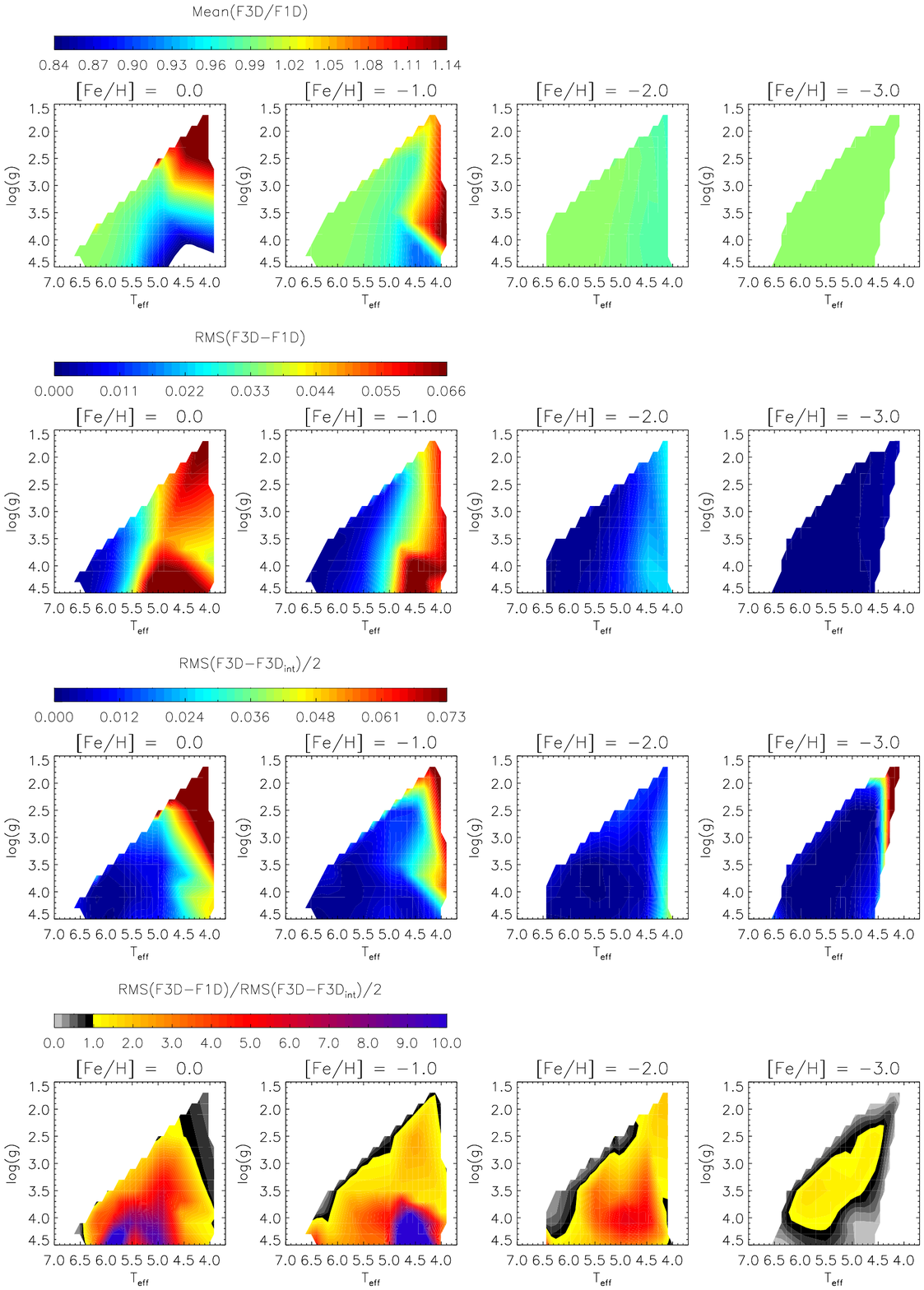}}
   \resizebox{1.5\columnwidth}{!}{\includegraphics[trim=0 410 0 40, clip]{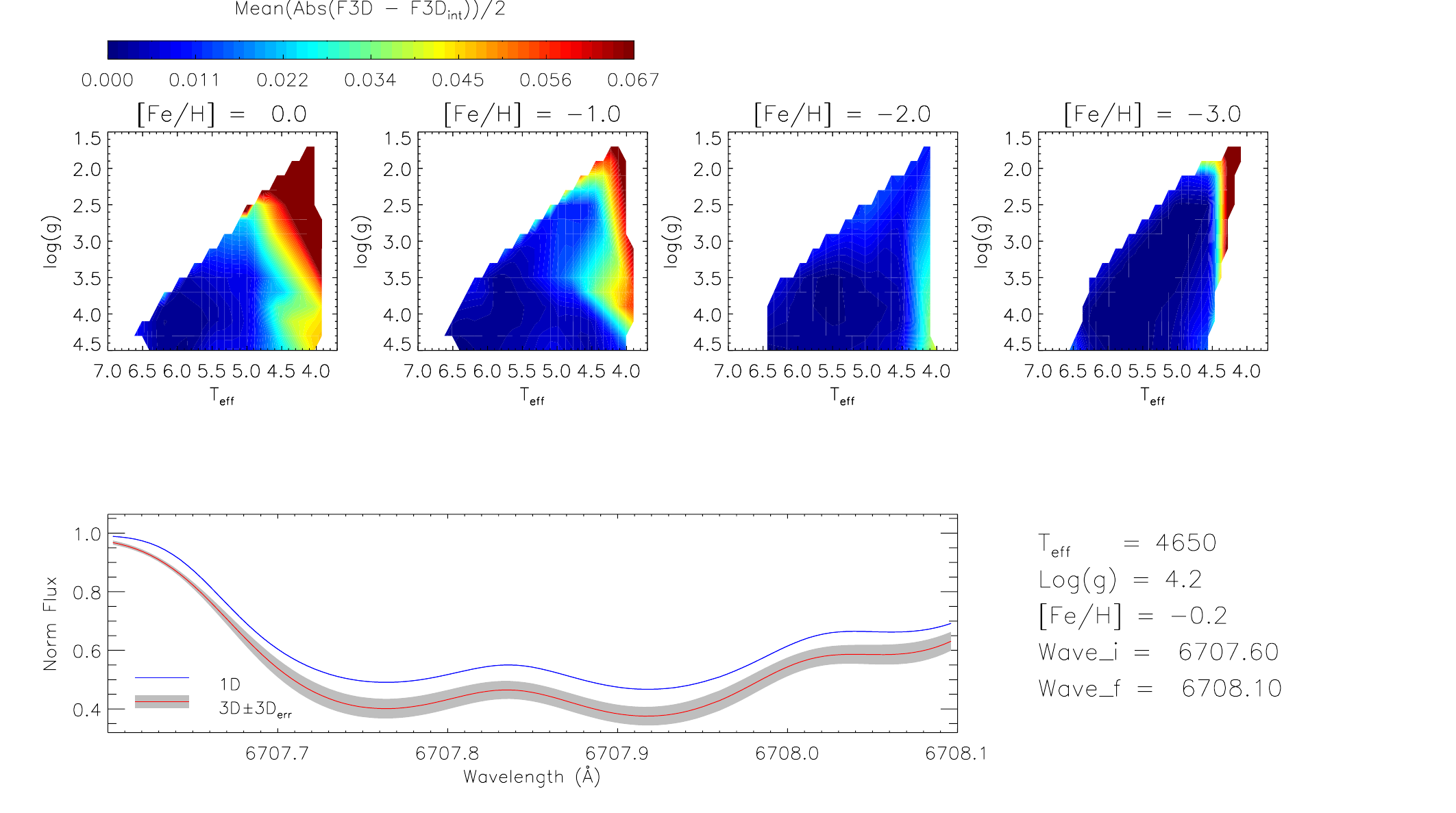}}
  \caption{Diagnostic plots for the Li resonance lines at 6707.6 and 6708.1~\AA. Effective temperature is expressed in $\times10^3$~K.}
   \label{fig:diagLi}
\end{figure*} 

The measurement of lithium abundance is relevant in many fields, such as the study of stellar interiors, primordial nucleosynthesis, stellar evolution, and exoplanet hosts. There are very few spectral lines available to measure its abundance, and the most popular one is the resonance doublet at $\sim$6707~\AA.

Because of the importance of lithium abundance, many studies have been devoted to exploring departures from LTE and 3D effects; for example   \citet{Carlsson94,Asplund99,Asplund03,Collet07,Wang21}. These studies concluded that both nonLTE and 3D effects are relevant for this transition, and its relative importance strongly depends on the stellar parameters. One-dimensional nonLTE corrections are largest for cool stars: at low Li abundances, overionization dominates and the correction is around +0.2~dex, while for high Li abundances, the line gets strengthened due to resonance scattering, and 1D nonLTE effects reach -0.6~dex for dwarfs and even more for giants \citep{Carlsson94}. When temperature inhomogeneities are taken into account in 3D models, the 1D LTE abundances in metal-poor subgiants and dwarfs with $T_{\rm eff}\sim$6000~K should be corrected by -0.2~dex. Corrections are more severe in red giants, reaching -0.5~dex.

The steep temperature variations in 3D model atmospheres cause departures from LTE larger than in 1D. Indeed, the large 3D LTE effects are in general compensated by 3D nonLTE corrections, which work in the opposite direction for the Li spectral line. In the case of the Sun, for example, nonLTE models reduce the EW of the line roughly by 20\%, which is close to the increase in 3D LTE \citep{Asplund05, Wang21}.

We explore 3D LTE effects in this line with our interpolation code, and compare our results with the literature. We select the wavelength range between 6707.6 and 6708.1~\AA, and the results are shown in Figure~\ref{fig:diagLi}. The most remarkable behavior of the 3D-1D differences arises at high metallicities and low temperatures, where a sharp dependence on gravity can be found. In this region of stellar parameters, the Li line has a much smaller EW in 1D than in 3D, while for giants it shows the opposite behavior. As we move towards [Fe/H]$<$-1, the severe changes with $\log g$ disappear, and 3D-1D differences diminish as the line loses strength. At [Fe/H]$=-2,$ the line is smaller in 1D as we move to low temperatures. At [Fe/H]$=-3,$ we do not see any difference between 3D and 1D, but a correction of $\bigtriangleup$(3D-1D)$=$-0.2dex at [Fe/H]=-3.6, $T_{\rm eff}\sim$6000~K, and $\log g\sim$4.5   has been reported \citep{Gonzalez08}. This is due to the fact that, at such low metallicities, the Li transition is very weak, and very small differences in EW can lead to appreciable changes in abundances.

According to our results, the 3D interpolation with our code is appropriate at almost all the stellar parameters studied, becoming poorer at low metallicities, where the errors of the interpolation are greater than the differences in fluxes. However, as highlighted above, nonLTE corrections are large for this resonance transition. As a resonance line, it originates in the outer atmosphere where 3D models predict a steep variation in temperatures, which make it more prone to departures from LTE.

Finally, we show the 1D and 3D spectra for the severe case of a dwarf with high metallicity and low temperature. As advanced by the contour plots, the 1D line is much smaller, and the abundances derived with a 1D LTE model would be overestimated compared to those of the 3D LTE model.

\subsection{H$\alpha$}

\begin{figure*}[]
   \centering
   \resizebox{1.5\columnwidth}{!}{\includegraphics[trim=0 35 0 20, clip]{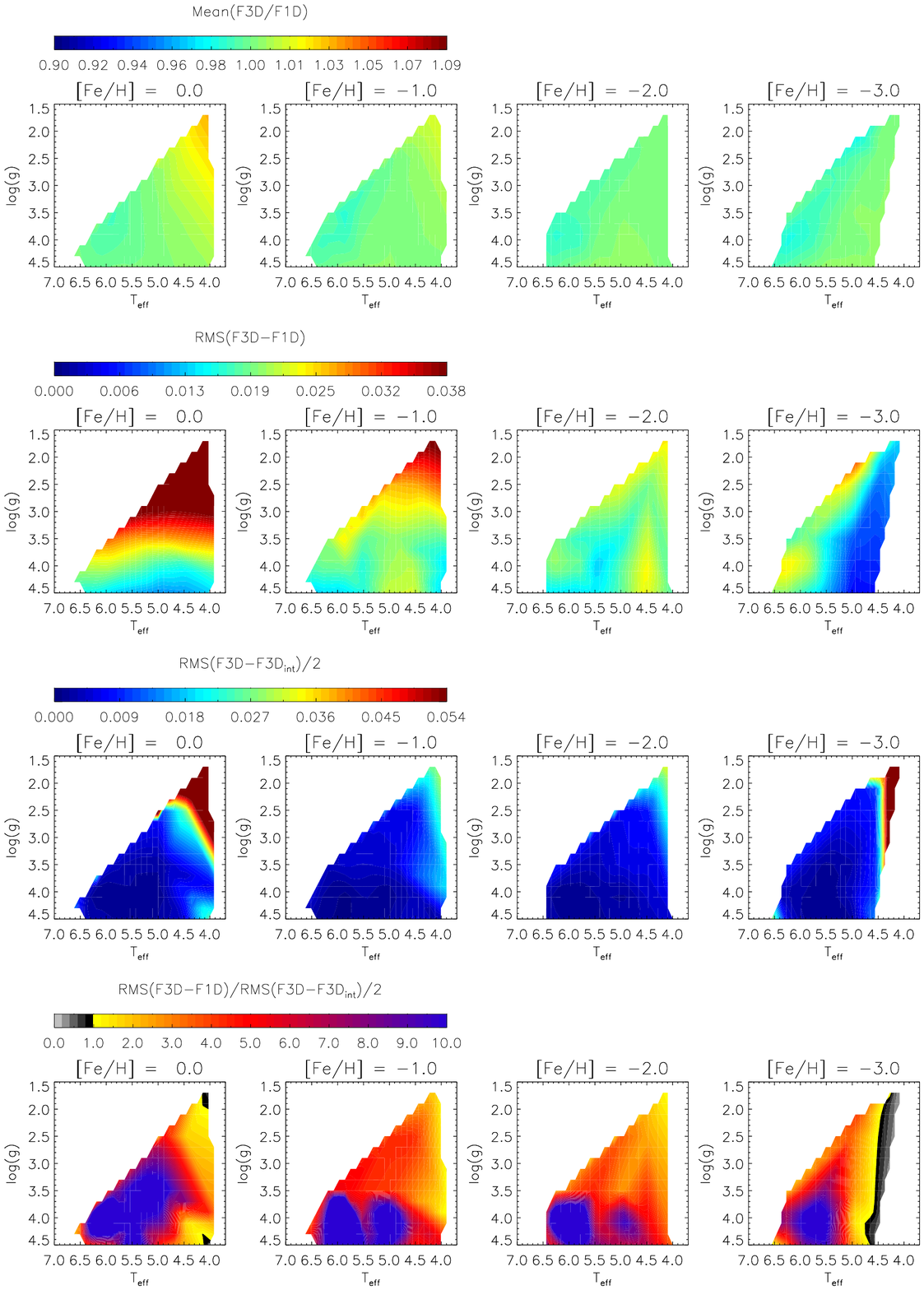}}
   \resizebox{1.5\columnwidth}{!}{\includegraphics[trim=0 410 0 40, clip]{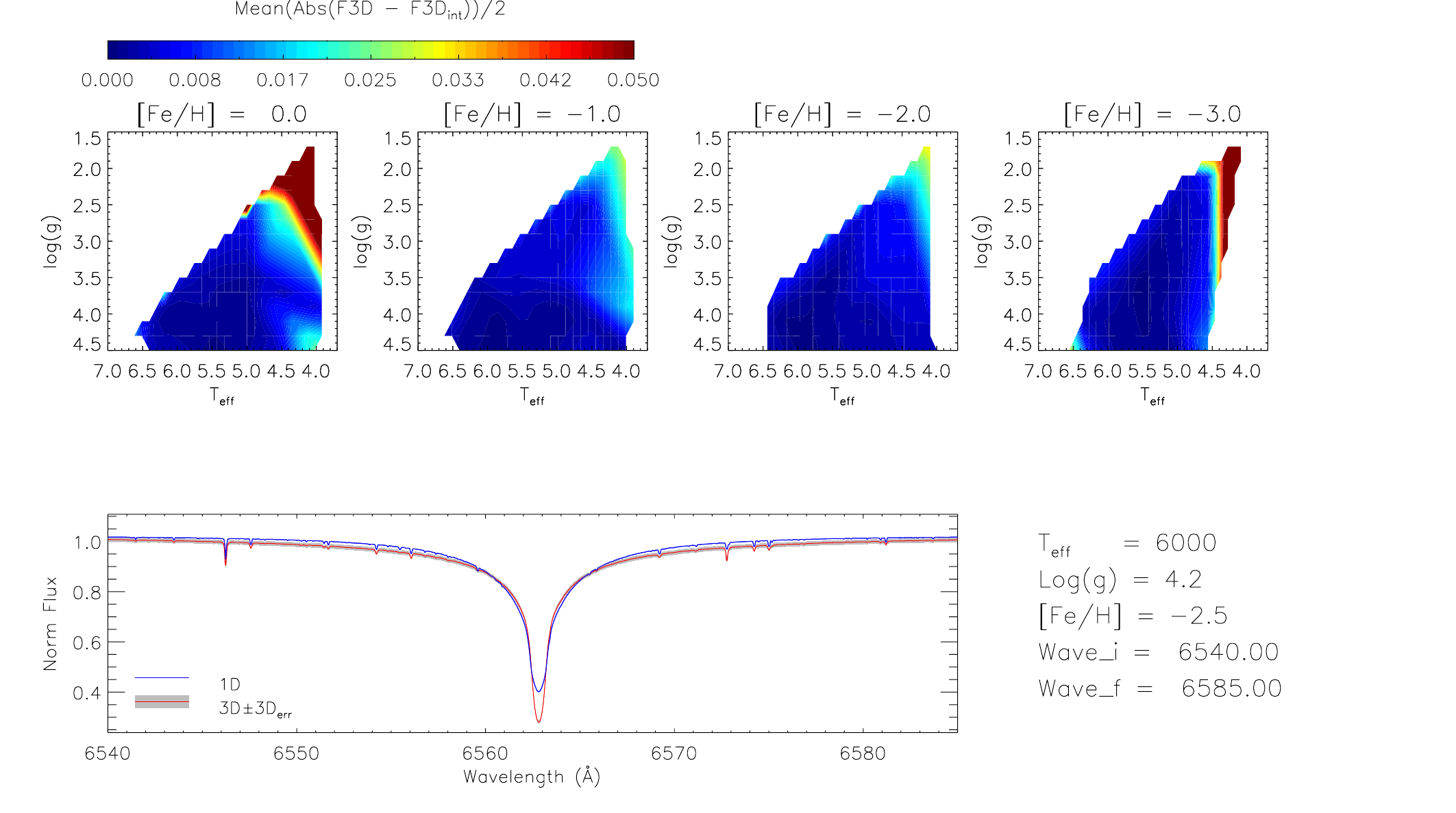}}
  \caption{Diagnostic plots for the H$\alpha$ line at 6562.72~\AA. Effective temperature is expressed in $\times10^3$~K.}
   \label{fig:diagHa}
\end{figure*}

Our last example is the H$\alpha$ line. This line is important because it can be used to determine effective temperatures, as the wings of the line are especially sensitive to small changes in the thermal structure of the atmosphere \citep[e.g.,][]{Cayrel88}. Our diagnostic plots do not show dramatic differences as in the case of lithium, but small differences can be observed, and in general the line would be slightly weaker in a 3D calculation than in its 1D counterpart.

The wavelength range that we have selected is wide enough to include several transitions in the wings of the H$\alpha$ line, and therefore one should be cautious when analyzing the results. At high metallicities and low temperatures, H$\alpha$ diminishes its strength, and the spectral features in the wings become more visible. The differences seen in this range of temperatures are probably dominated by these other lines rather than by the H$\alpha$ line itself. On the other hand, the core of the line can be misrepresented for giants in 1D because the microturbulence for these spectra was selected to be 1~km~s$^{-1}$ for all stars.

Besides the possible interferences mentioned, we can still detect differences between 3D and 1D interpolations that reach $\sim$5\% in flux in giant stars, and $\sim$3\% for dwarfs. In general, errors in the 3D interpolation are small, and therefore the interpolation  code is providing useful results.

Finally, in the bottom panel of Figure~\ref{fig:diagHa} we show H$\alpha$ for a dwarf star at low metallicity and high temperature. As can be seen, the core of the line is much more intense in 3D. The wings of the line are also different, and the ratio between 3D and 1D changes as we move through the wing. Close to the core, the 3D spectrum is narrower, and at $\sim$2.5~\AA~ from the core this behavior is inverted. Therefore, if we use the inner part of the wings to fit the temperature with a 1D model, we will be overestimating the effective temperature, because the line is wider for higher temperatures. On the contrary, if we fit the wings further than 2.5~\AA~ from the core with a 1D model, we underestimate the effective temperature compared to the 3D result.

NonLTE effects may play an important role in the H$\alpha$
line. \citet{Przybilla04} studied the behavior of 1D LTE models compared to that of 1D nonLTE
models for the H$\alpha$ line in the Sun, and found more intense line cores in the nonLTE case. For the solar case, we do not find any difference
between 3D and 1D spectra regarding the core of the line. However, nonLTE
corrections in 1D cannot be extrapolated to 3D. \citet{Leenaarts12} studied
the same line for the Sun with 3D nonLTE models and also found a more intense
core compared to our 3D LTE models. A quantitative analysis would be necessary
to draw further conclusions about nonLTE corrections to 3D models for the
H$\alpha$ spectral line.

A word of caution about the core of H$\alpha$ and other very intense lines is necessary,  because they are most likely formed in layers where models become unrealistic because of 
missing physics such as magnetic fields and unresolved shocks.

\section{Results and Conclusions}

We present an algorithm and a tool developed to interpolate spectra from the CIFIST grid of 3D hydrodynamical simulations of stellar atmospheres. The implementation of the interpolation algorithm is carried out using IDL,
and we describe its workflow  here in detail.  As input, the code takes a
wavelength range, a set of stellar parameters (the interpolation point), and
optionally a 1D spectrum together with its spectral resolution. If a 1D spectrum is provided, the code interpolates the ratio 3D/1D of normalized fluxes, which yields lower errors. Otherwise, a straight interpolation of the 3D grid is performed. The code returns three arrays with wavelength, interpolated and normalized 3D spectra, and interpolated flux errors. In addition, several diagnostic plots are provided at request that can be used to   quantitatively analyze the importance of the 3D effects, the quality of the interpolation in terms of errors, and the improvement of the 3D interpolated spectrum over a 1D spectrum.

We selected some spectral regions of interest to show examples of applications of the interpolation code, which are also useful to further explain how the code is used and its limitations. In particular, we demonstrate the performance of the  code through the diagnostic plots of the IR Ca I triplet, the CO molecular band at 16611\AA, the Li resonance, and the H$\alpha$ lines.

The expected errors in the interpolated flux are in general under 2\%, which allows the user to check whether or not 3D abundance correction may be important in practical applications. Nevertheless, regions like the Balmer discontinuity may not be as reliable, and in general errors are highly dependent on the wavelength range and the target stellar parameters. More accurate predictions are clearly desirable, but we deem the strategy described in this paper ---based on flux ratios and interpolation with radial basis functions--- to be a good first step toward making the predictions from 3D models accessible to researchers. We  encourage users not to make blind use of the code: we strongly recommend examining the output plots prior to systematic use of the code.
Our tests demonstrate the potential of the code in identifying differences between predicted spectra from 3D and 1D LTE model atmospheres, and the overall small errors in the interpolated spectrum. 

Finally, it is of utmost importance to stress that, sometimes, 3D LTE corrections are compensated for by nonLTE effects. Therefore, caution should be exercised because all the calculations in the CIFIST grid have been performed assuming LTE. When departures from LTE are important, they can be quite different in 1D and 3D, and the differences between the two can easily be much larger than those we find in our calculations. In addition, we consider flat values for the micro- and macroturbulence velocities in our 1D calculations, and ignore stellar rotation, which hampers   comparisons with observations.

\begin{acknowledgements}

The calculations to obtain the original grid of 3D and 1D spectra were performed at the Texas Advanced Computing Center (TACC). This research made use of computing time available on the high-performance computing systems at the Instituto de Astrofisica de Canarias. The authors  thankfully acknowledge the technical expertise and assistance provided by the Spanish Supercomputing Network (Red Espanola de Supercomputacion), as well as the computer resources used: the LaPalma Supercomputer, located at the Instituto de Astrofisica de Canarias.

S.B. and C.A.P. acknowledge financial support from the Spanish Government through the 
projects AYA2014-56359-P and AYA2017-86389-P. H.-G.L. acknowledges financial support by the Sonderforschungsbereich SFB\,881
``The Milky Way System'' (subprojects A4) of the German Research Foundation
(DFG). S.B. also thanks Dr. Andr\'es del Pino for his support and help during this project.

\end{acknowledgements}

\nocite{*}

\bibliographystyle{aa} 
\bibliography{3dcode_v6} 

\end{document}